# SIGHT: A Methodological Appraisal Framework for Evaluating Large Language Models in Ophthalmology

Hadi Vahedi [1], Fatemeh Razmjooei [2], Farbod Semnani[3], Parviz Vahedi [4], Roohallah Alizadehsani [5, *], Ali Jafarizadeh [6, 7, *]

[1] Research Center for Evidence-Based Health Management, Maragheh University of Medical Sciences, Maragheh, Iran

[2] Poostchi Ophthalmology Research Center, Department of Ophthalmology, Medical School, Shiraz University of Medical Sciences, Shiraz, Iran

[3] School of Medicine, Tehran University of Medical Sciences, Tehran, Iran

[4] Department of Anatomical Sciences, Maragheh University of Medical Sciences, Maragheh, Iran

[5] School of Medicine, Queensland University of Technology, Brisbane, Australia

[6] Research Center for Evidence-Based Medicine, Faculty of Medicine, Tabriz University of Medical Sciences, Tabriz, Iran

[7] Nikookari Eye Center, Tabriz University of Medical Sciences, Tabriz, Iran.

* Corresponding authors

1. Roohallah Alizadehsani, PhD

   School of Medicine, Queensland University of Technology, Brisbane, Australia

   Tel: +61 3 524 79394

   Email: roohallah.alizadehsani@qut.edu.au

   ORCiD: https://orcid.org/0000-0003-0898-5054

2. Ali Jafarizadeh, MD, MPH

   Research Center for Evidence-Based Medicine, Iranian EBM center: A JBI Centre of excellence, Faculty of Medicine, Tabriz University of Medical Sciences, Tabriz, Iran

Nikookari Eye Center, Tabriz University of Medical Sciences, Tabriz, Iran.

Email: Jafarizadeha@tbzmed.ac.ir , Ali.jafarizadeh.md@gmail.com

ORCiD: https://orcid.org/0000-0003-4922-1923

| | | | |
|---|---|---|---|
| Hadi Vahedi | MD | hadivahedimrg@gmail.com | 0009-0008-4769-451X |
| Fatemeh Razmjooei | MD, HMBA | Razftmh@gmail.com | 0009-0000-3757-5172 |
| Farbod Semnani | MD, MPH | farsem12@gmail.com | 0000-0002-8964-419X |
| Parviz Vahedi | PhD | pa.vahedi48@gmail.com<br>Parviz_vahedi48@mrgums.ac.ir | 0000-0003-4850-8519 |
| Roohallah Alizadehsani | PhD | roohallah.alizadehsani@qut.edu.au | 0000-0003-0898-5054 |
| Ali Jafarizadeh | MD, MPH | Ali.jafarizadeh.md@gmail.com;<br>Jafarizadeha@tbzmed.ac.ir | 0000-0003-4922-1923 |

**Abstract**

Large language models (LLMs) are increasingly evaluated across ophthalmic tasks, yet the methodological quality of these evaluations has received little attention. Most studies report accuracy while overlooking design features required for valid and clinically meaningful inference. We critically reviewed published ophthalmic LLM evaluation studies covering knowledge assessment, patient education, diagnostic reasoning, triage, clinical documentation, multimodal image interpretation, and uncertainty estimation. We used the methodological weaknesses that recurred across study designs to develop the SIGHT framework, which has five domains: Sample and item provenance, Independence and blinding, Ground truth and comparator fairness, Handling of paired and repeated-measures data, and Transparency and trustworthiness. Current evaluations rely heavily on benchmark accuracy, synthetic clinical vignettes, and curated datasets, and real-world clinical validation remains rare. Common limitations include potential training-data contamination, inadequate handling of paired data, limited assessment of inter-rater reliability, inconsistent reference standards, insufficient evaluation of calibration and uncertainty, and frequent overinterpretation of benchmark performance as evidence of clinical readiness. SIGHT provides twenty structured appraisal items to evaluate internal validity, interpretability, and applicability across diverse ophthalmic LLM study designs. It offers a practical way to appraise ophthalmic LLM studies and may improve the methodological quality, interpretability, and clinical relevance of future evaluations.

## 1. Introduction

Ophthalmology adopted artificial intelligence (AI) early and with unusual confidence. Convolutional neural networks for diabetic retinopathy screening moved from proof-of-concept to regulatory clearance within a few years (Abràmoff et al., 2018; Gulshan et al., 2016; Ipp et al., 2021), and the specialty developed, comparatively quickly, an evidentiary culture around image-based AI: labelled datasets(De Fauw et al., 2018), external validation cohorts (Lee et al., 2021; Ting et al., 2017), prospective screening trials (Ruamviboonsuk et al., 2022; Wolf et al., 2024), and eventually regulatory pathways. That culture has not transferred to large language models (LLMs).

Since late 2022, general-purpose LLMs have been evaluated against board examinations, patient questions, clinical vignettes, referral letters, operative notes, imaging cases and educational materials(Kung et al., 2023; Lee et al., 2023; Singhal et al., 2023; Thirunavukarasu et al., 2023). The volume is remarkable. One systematic review identified 187 quantitative ophthalmic LLM evaluations published up to November 2024 (Zhang et al., 2025); a broader medical review counted 519 LLM evaluations across all of medicine to February 2024, of which only 5% used real patient care data (Bedi et al., 2025). Bibliometric analysis of first-quartile ophthalmology journals alone identified 101 eligible articles by mid-2024, of which roughly one third were original investigations (Agnihotri et al., 2025).

This literature has been repeatedly mapped. Zhang et al. classified ophthalmic LLM evaluations into six domains, medical query answering, standardized examinations, diagnostic decision-making, research assistance, patient triaging, and disease prediction (Zhang et al., 2025). Scoping and systematic reviews now exist for glaucoma, for primary-care ophthalmic education, and for vision and vision-language foundation models (Jin et al., 2025; Li et al., 2026; Rubegni et al., 2026). A reader arriving at this field today is not short of maps(Artiaga et al., 2025).

What the field lacks is a critical reading. The existing syntheses answer the question what has been studied? with considerable rigor. They do not answer the question that matters for clinical adoption: given how these studies were designed, what are we actually entitled to believe(Agrawal et al., 2025)?

An exploratory meta-analysis illustrates the problem. Despite reporting a pooled diagnostic accuracy of 59.4%, heterogeneity was extremely high ($I^2$ = 94.5%), and subgroup analysis unexpectedly suggested that GPT-3.5 outperformed GPT-4 (Zhang et al., 2025). This unlikely finding is best interpreted as evidence of methodological inconsistency rather than a true difference in model performance. In other words, variation in how studies are conducted currently exceeds the signal we are trying to measure, highlighting the need for better study design and standardized evaluation rather than simply more studies.

This review therefore takes a different organizing principle from its predecessors. We survey performance evidence domain by domain, as others have. But at each domain we ask what design features would be required to support the claims being made, and whether they are present. We then consolidate these observations into a five-domain appraisal instrument (the SIGHT framework) intended for use by investigators designing studies, by peer reviewers assessing them, and by future systematic reviewers weighting them.

Our central argument is that evaluation design may be an increasingly important bottleneck in translating LLMs into ophthalmic care, as model capabilities continue to evolve rapidly, while evaluation methodologies may not have kept pace with these advances. The methodological challenges addressed in this work, together with the proposed SIGHT framework and its intended users, are summarized in the Figure 1.

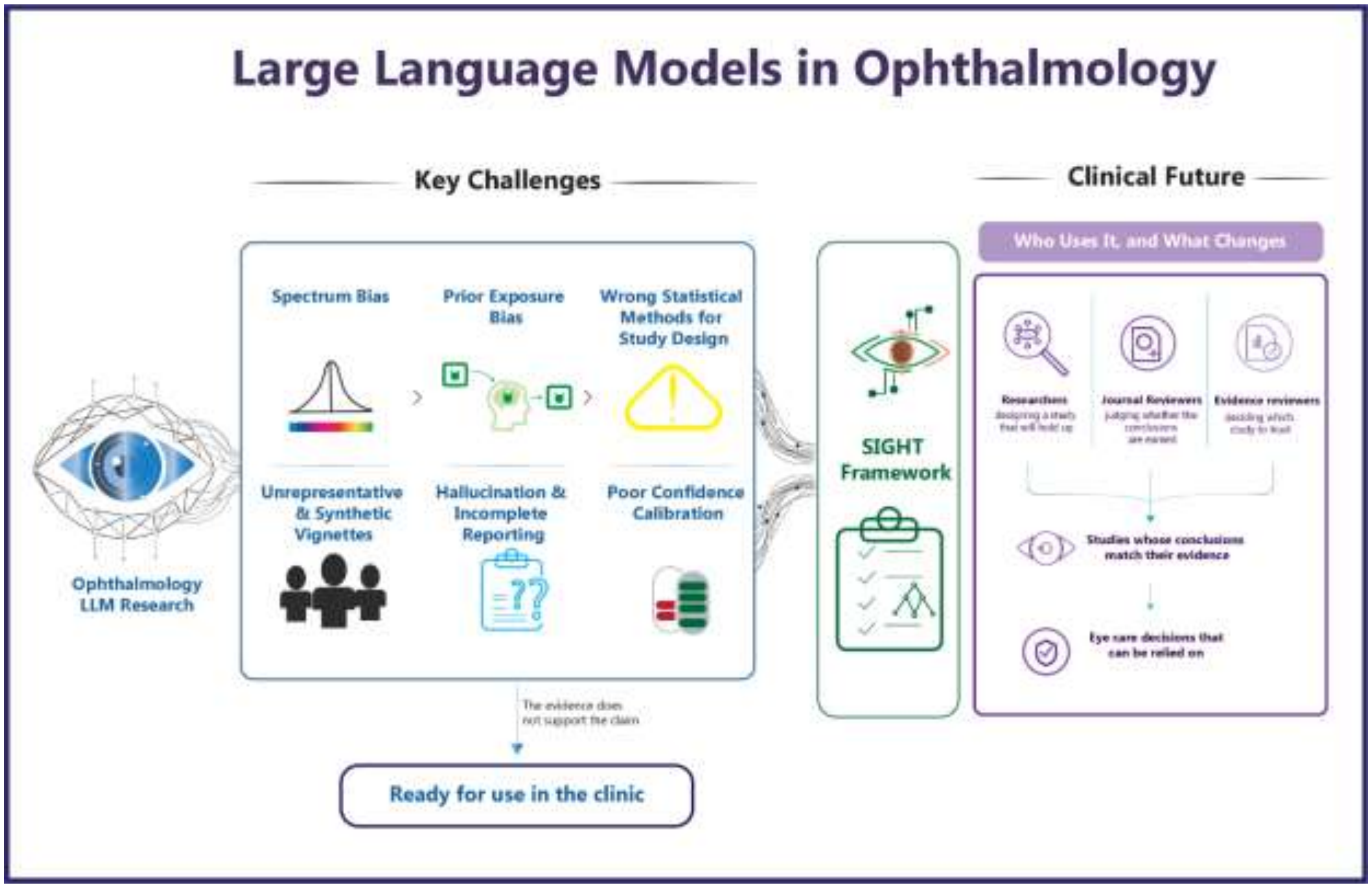

**Fig. 1.** Overview of the methodological challenges in ophthalmology large language model (LLM) research and the proposed SIGHT framework (Sample and item provenance, Independence and blinding, Ground truth and comparator fairness, Handling of paired and repeated-measures data, and Transparency and trustworthiness). Current studies are commonly affected by spectrum bias, prior exposure bias, statistical methods inappropriate for the study design, unrepresentative and synthetic vignettes, hallucination and incomplete reporting, and poor confidence calibration, so the available evidence does not support the claim that these models are ready for clinical use. The SIGHT framework addresses these limitations and is intended for researchers designing a study, journal reviewers judging whether conclusions are warranted, and evidence reviewers deciding

which studies to trust, leading to studies whose conclusions match their evidence and, ultimately, to eye care decisions that can be relied on.

## 2. Methods

### 2.1 Review design

This is a structured narrative review, assessed using with the Scale for the Assessment of Narrative Review Articles (SANRA) quality criteria (Baethge et al., 2019). A narrative rather than systematic format was chosen deliberately: three systematic reviews and one meta-analysis of this literature already exist, and the marginal value of a fourth enumeration is low (Li et al., 2026; Rubegni et al., 2026; Wu et al., 2024; Zhang et al., 2025). The contribution intended here is interpretive and methodological, which a narrative structure accommodates and a structure constrained by the Preferred Reporting Items for Systematic Reviews and Meta-Analyses (PRISMA) statement does not (Greenhalgh et al., 2018).

### 2.2 Search strategy

We searched PubMed, Scopus, Web of Science and Google Scholar for English-language literature published between January 2022 and July 2026, combining terms for large language models ("large language model", "LLM", "ChatGPT", "GPT-4", "Gemini", "Claude", "generative artificial intelligence", "chatbot", "foundation model") with terms for ophthalmology and its subspecialties. Reference lists of identified reviews were hand-searched, and forward citation chaining was performed on the four principal syntheses (Li et al., 2026; Rubegni et al., 2026; Wu et al., 2024; Zhang et al., 2025).

We supplemented this with targeted searches of the methodological literature on LLM evaluation, including reporting guidelines registered with the Enhancing the QUAlity and Transparency Of health Research (EQUATOR) Network (Collaborative, 2025; Collins et al., 2024; Gallifant et al., 2025; Liu et al., 2020; Vasey et al., 2022).

### 2.3 Development of the SIGHT appraisal framework

Until recently, no risk-of-bias instrument existed for medical LLM evaluation studies, an absence explicitly acknowledged by the largest systematic review in this space, which was obliged to weight all included studies equally irrespective of methodological quality (Zhang et al., 2025). That gap has now been partly filled. The Alberta Quality Assessment Tool: Risk of Bias (AQAT:RoB), developed through modified Delphi with 19 panellists and pilot-validated on 16 studies (86.1% agreement, Cohen κ 0.70), assesses risk of bias in medical LLM question-answer studies across five domains: Questions, Reference Answers, LLM Answers, Evaluators, and Outcomes (Ye et al., 2026).

AQAT:RoB is the appropriate foundation, and SIGHT is constructed to complement rather than duplicate it. Four gaps define the space SIGHT occupies, none of them specialty-specific.

**Eligible study designs.** AQAT:RoB applies to question-answer studies "that involve human participants in assessing the outputs of AI models." This excludes automated-scoring benchmark studies (the dominant design in standardized examination research) as well as multimodal image interpretation and documentation-generation tasks, none of which necessarily involve human rating of free-text output. As a result, at least 41 of the 187 studies (21.9%) identified in the largest systematic review of ophthalmic LLMs fall outside the types of studies AQAT:RoB is designed to evaluate (Zhang et al., 2025).

**Statistical validity.** AQAT:RoB contains no items addressing analysis. The paired and repeated-measures structure that characterizes almost every multi-model comparison is therefore unassessed, as are multiplicity and effect-size reporting. A study can be judged low risk of bias throughout while having analyzed paired data with an independent-samples test.

**Reliability signals.** Neither calibration (whether a model's confidence tracks its correctness) nor the faithfulness of model-stated reasoning is covered, and model version is not requested despite undisclosed snapshot revisions materially affecting reproducibility.

**Concordance between evidence and claims.** Risk-of-bias instruments assess whether a study's methods could have produced a biased result. They do not assess whether the conclusions drawn exceed what the methods can support, and in this literature that is where much of the damage occurs. A methodologically clean benchmark study that reports accuracy and concludes the model is "comparable to ophthalmologists" or "ready for clinical use" has committed no bias in AQAT:RoB's sense, yet its central claim is unsupported. Several SIGHT items are therefore conditional on claim scope rather than on design alone: a study confining its conclusions to what it measured is marked not applicable, while one making deployment, human-comparability, or interpretability claims is assessed against the evidence those claims require. We therefore chose to address this explicitly. In the current LLM literature, unsupported conclusions are often more consequential than methodological flaws themselves, because they directly influence perceptions of clinical readiness and safety(Dong et al., 2026). A model evaluated only on benchmark tasks may nevertheless be presented as comparable to clinicians or suitable for deployment, encouraging inappropriate translation into practice. SIGHT therefore evaluates not only methodological validity, but also whether a study's conclusions are justified by its evidence.

SIGHT therefore extends appraisal of LLM evaluation studies to designs and dimensions that existing instruments do not reach. Its items are general; we develop and illustrate them in ophthalmology because that literature is unusually mature, has been systematically catalogued, and spans the full range of task types from examination benchmarking to multimodal interpretation.

The resulting framework comprises five domains and twenty items (Figure 2); each item is specified in full in Section 4.

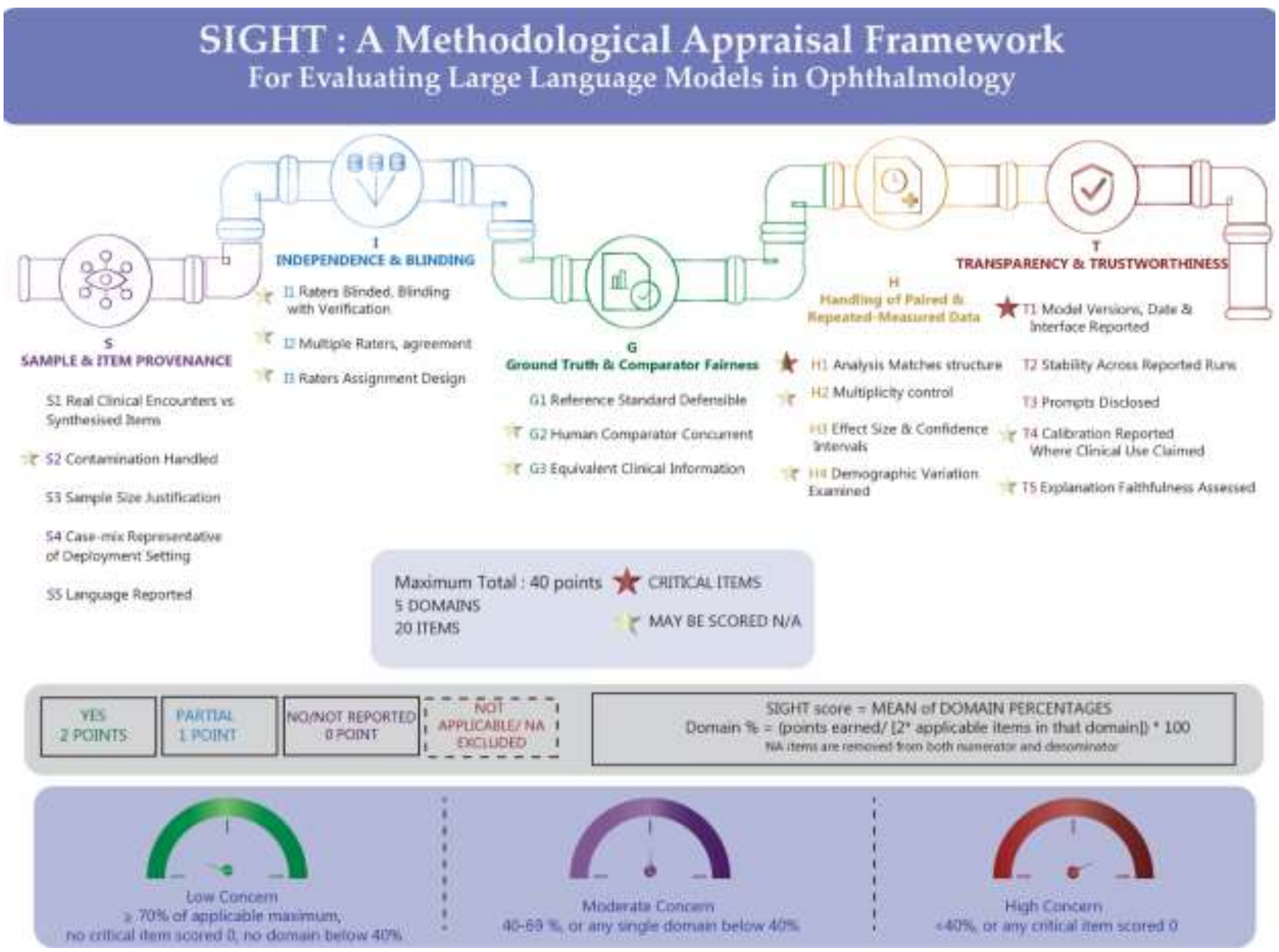

**Fig. 2.** The SIGHT framework for methodological appraisal of large language model evaluation studies. Five domains contain twenty items, scored Yes (2), Partial (1), or No/Not reported (0).

## 3. The evidence, domain by domain

The current ophthalmic LLM literature spans a wide range of clinical applications, from knowledge examinations and patient education to diagnosis, triage, documentation, and multimodal image interpretation. For each domain, we summarize the available evidence and then examine the methodological issues that determine how confidently those findings can be interpreted. The emphasis is therefore not only on what LLMs can do, but also on how reliably current studies demonstrate those capabilities.

Figure 3 summarizes the distribution of reported outcomes and task domains across the 187-study corpus that this section examines.

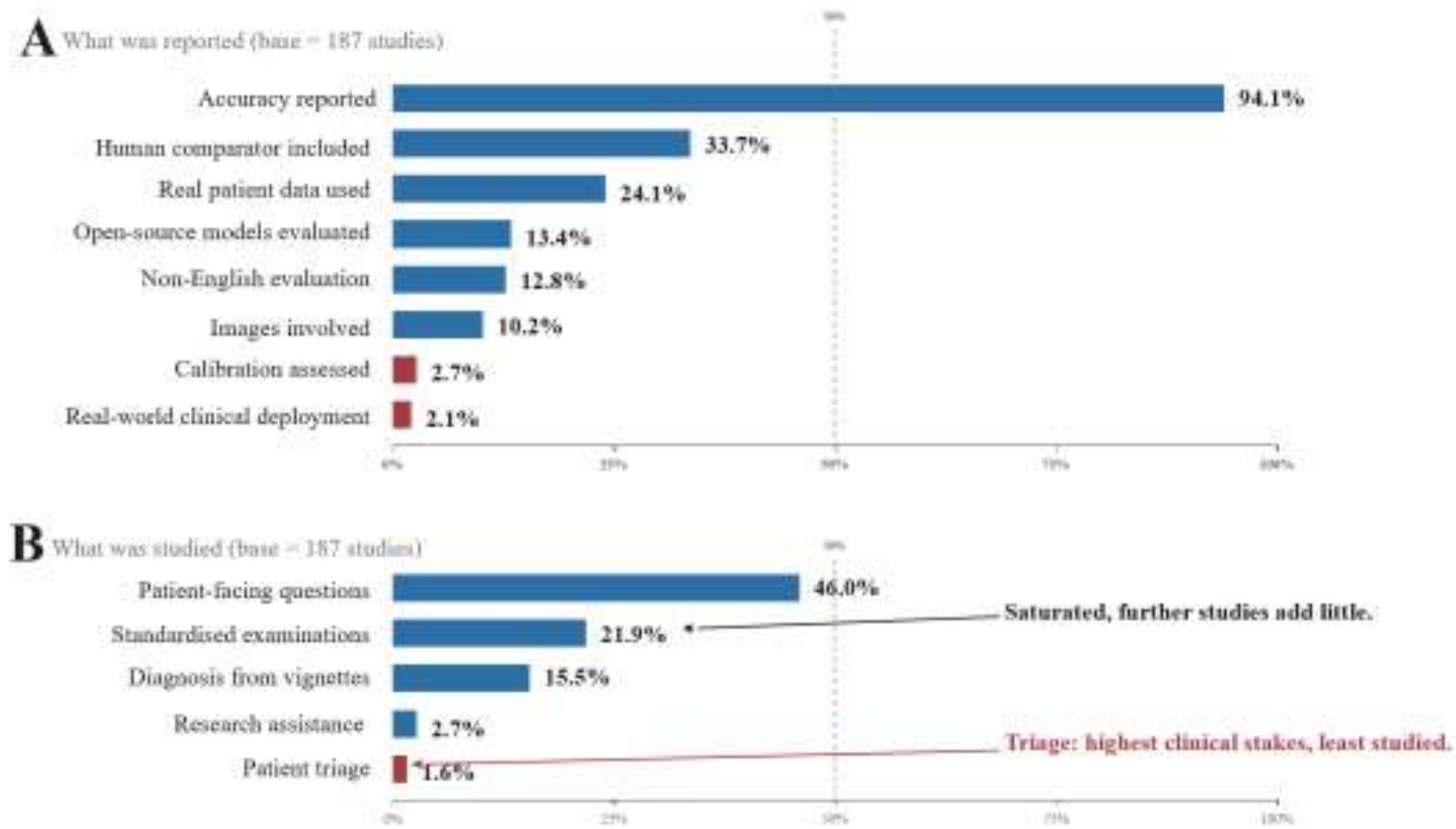

**Fig. 3.** Distribution of reported outcomes (A) and task domains (B) across 187 ophthalmic large language model evaluation studies. Accuracy was reported in 94.1% of studies; calibration (whether a model's confidence corresponds to its correctness) was assessed in 2.7%, and only 2.1% involved real-world clinical deployment. Data from Zhang et al. (Zhang et al., 2025).

### 3.1 Standardized examinations and ophthalmic knowledge assessment

This is the most mature and most saturated domain in the field, and the one where further primary studies now add least(Antaki et al., 2023; Haddad & Saade, 2024; Lin et al., 2023; Raimondi et al., 2023; Taloni et al., 2023; Thirunavukarasu et al., 2024).

The trajectory is well documented. Early evaluations of GPT-3.5 against ophthalmology board practice banks were unimpressive; Mihalache and colleagues found the model answered fewer than half of items correctly on a widely used preparation resource, with a re-analysis one month later showing improvement to 58% (Mihalache et al., 2023). Cai et al. compared three models against 250 Basic Science and Clinical Science Self-Assessment Program items and against human respondents: humans averaged 72.2%, GPT-3.5 scored 58.8%, and GPT-4.0 (71.6%) and Bing Chat (71.2%) performed comparably to humans (Cai et al., 2023). Antaki et al. reported GPT-4 exceeding 70% on simulated board-style examinations, not significantly higher than humans (Antaki et al., 2024), and later work documented incremental gains on Ophthalmic Knowledge Assessment Program (OKAP)-style question banks  (Teebagy et al., 2023).

Wu et al. meta-analyzed fourteen such studies and reported pooled accuracy of 0.65 (95% confidence interval [CI] 0.61–0.69), with GPT-4 and Bing Chat the top performers (Wu et al., 2024). Their conclusion that performance is "acceptable but not exceptional" has held up.

Two findings from this domain deserve to be carried forward rather than re-established:

**Subspecialty variance is large and consistent in direction.** Bard answered 62.4% of 250 BoardVitals items correctly overall, but 84% of oculoplastic items and only 24% of retina and vitreous items (Botross et al., 2024). This pattern (strong on descriptive, anatomically anchored subspecialties, weak on retina) recurs across studies and models and is unlikely to be an artefact of any single question bank (Galindo et al., 2026).

**Multiple-choice performance overstates competence.** The most methodologically informative recent study in this domain assessed both formats. Against ophthalmology continuing medical education questions, GPT-4 achieved 82.5% (99/120) on multiple choice and Gemini Pro 1.5 49.2% (59/120). When the same models produced free-text prose answers assessed by three board-certified ophthalmologists, the picture changed sharply: raters identified incorrect reasoning in 42% and 58% of responses, inappropriate content in 29% and 36%, missing content in 42% and 30%, and possibility of physical or emotional harm in 36% and 44% (Chen et al., 2026).

This study highlights a key limitation of the examination literature: strong performance on multiple-choice questions does not necessarily translate into safe clinical reasoning. A model may choose the correct answer yet provide incomplete, incorrect, or potentially harmful explanations when asked to justify its decision. Therefore, board-style examination scores should be viewed as an indirect measure of clinical competence rather than evidence of safe real-world performance. Figure 4 shows this divergence directly: the same models and the same questions, with only the response format changed.

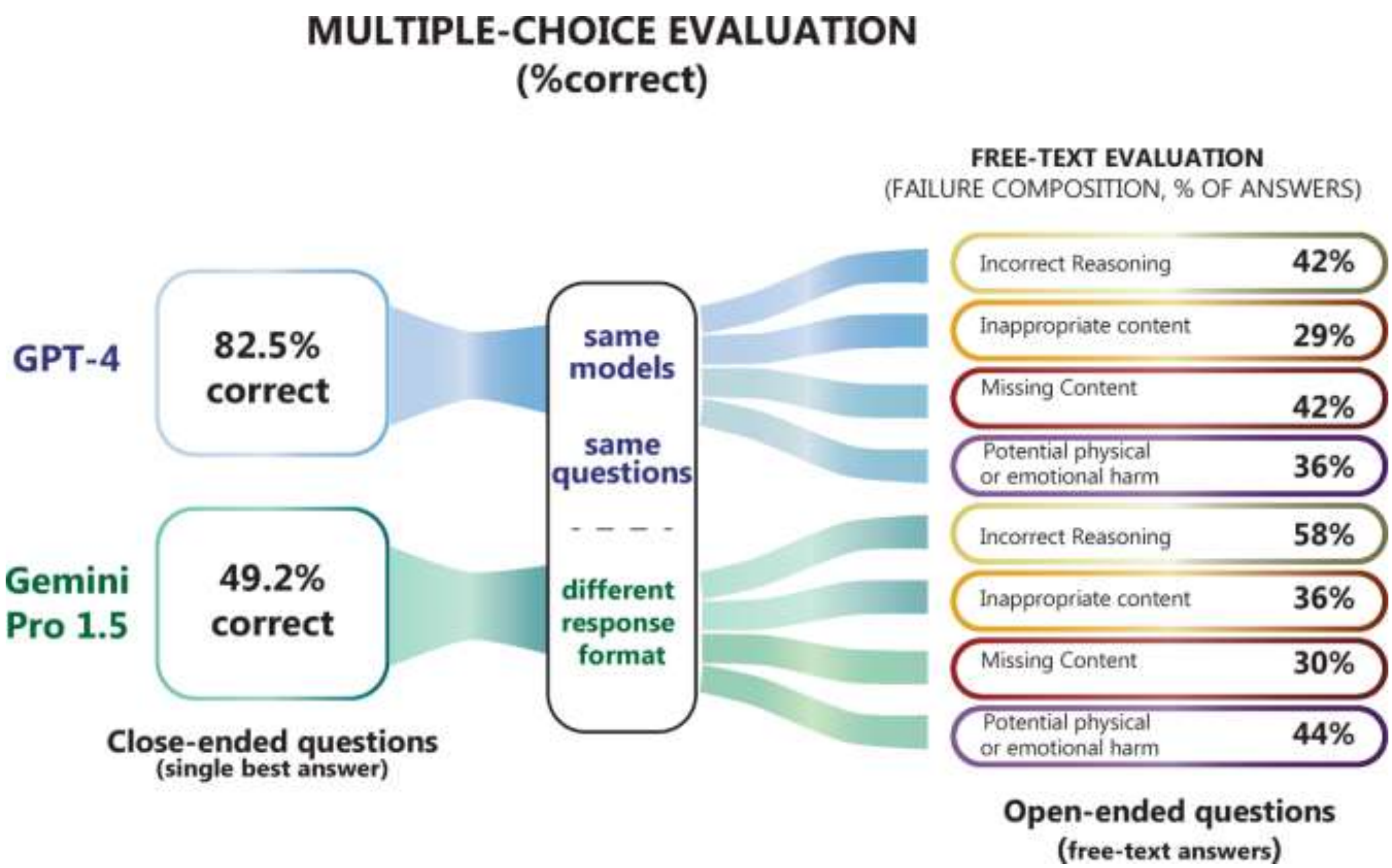

**Fig. 4.** Multiple-choice accuracy does not predict the safety of the same model's free-text answers. Two models answered the same set of ophthalmic continuing medical education questions in two

response formats. In the closed-ended format, requiring selection of a single best answer, GPT-4 answered 82.5 percent correctly and Gemini Pro 1.5 answered 49.2 percent correctly. When the same models produced open-ended free-text answers to the same questions, three board-certified ophthalmologists identified incorrect reasoning in 42 percent and 58 percent of answers, inappropriate content in 29 percent and 36 percent, missing content in 42 percent and 30 percent, and the possibility of physical or emotional harm in 36 percent and 44 percent. Failure categories are not mutually exclusive, so percentages do not sum to 100. Data from Chen et al. (Chen et al., 2026).

**Appraisal note. Benchmark contamination.** Almost no study in this domain addresses whether the item bank used appeared in model training data. Many commercial and educational question banks have substantial overlap with content available on the open web; Basic and Clinical Science Course (BCSC)-derived, OKAP-derived and StatPearls-derived items are plausibly present in pre-training corpora. Contamination would inflate accuracy without any corresponding gain in reasoning ability, and would do so differentially across model generations with different training cutoffs. Contamination-resistant designs exist (newly written items, held-out items, perturbation of surface form with preserved clinical content) and are essentially unused. We regard this as one of the most important unexamined threats to validity in the examination literature (C. Deng et al., 2024; Golchin & Surdeanu, 2024; Sainz et al., 2023) .

### 3.2 Patient-facing question answering and patient education

This is the largest task category in the field, accounting for roughly 46% of studies in the Zhang et al. (Zhang et al., 2025), and the domain with the clearest signal of genuine clinical relevance(Ayers et al., 2023; Cappellani et al., 2024; Tailor et al., 2024; Taloni et al., 2025).

The clinical argument for LLMs in this domain rests substantially on access: a global shortage of ophthalmologists, unevenly distributed, against rising demand from ageing populations(Burton et al., 2021; Resnikoff et al., 2012; Resnikoff et al., 2020; Steinmetz et al., 2021). If that is the case for deployment, then performance in low-resource settings and non-dominant languages is not a secondary consideration, it is the primary outcome.

Bernstein et al. found that chatbot responses to online patient eye care questions were broadly comparable in quality to those written by ophthalmologists (Bernstein et al., 2023). Subsequent work has replicated this across subspecialties: myopia care, conjunctivitis, dry eye disease in Chinese-language patients, age-related macular degeneration, childhood glaucoma education, optic disc drusen, orbital and oculofacial disorders, and lacrimal drainage disorders, and even ocular symptom queries or surgical treatments for retinal disease (Ali, 2023; Balas et al., 2024; Dihan et al., 2024; Ferro Desideri et al., 2023; Lim et al., 2023; Momenaei et al., 2023; Potapenko et al., 2023; Pushpanathan et al., 2023; Shi et al., 2024; Wang et al., 2025).

Three robust findings emerge.

**Content quality is generally adequate; readability is generally not.** ChatGPT-generated responses to common ophthalmology questions cluster around a tenth-grade reading level, and ChatGPT-4o responses to real patient portal messages averaged a Flesch–Kincaid grade level of 12.1 ± 2.1 (Hamzeh et al., 2026; Riazi Esfahani et al., 2025). Both sit far above the sixth- to seventh-grade level recommended for patient education materials(Edmunds et al., 2013; Williams et al., 2016). LLM outputs are consistently less readable than professionally produced pamphlets even where they are more comprehensive.

**Language disparity.** Non-English evaluation appeared in 12.8% of ophthalmic LLM studies, and of those, Chinese, Spanish and Japanese accounted for the large majority; the remaining languages appeared once each(Zhang et al., 2025). Where language disparity has been examined directly, it is substantial: performance on retinal vascular disease classification differs materially between English and Chinese prompts, and multilingual ophthalmic question-answering shows systematic degradation outside high-resource languages (Jin et al., 2024; Liu et al., 2024; P. Qiu et al., 2024; Restrepo et al., 2024; Sakai et al., 2023).

**Prompt-based readability optimization appears feasible without substantially compromising content quality.** In a study, instructing ChatGPT-4o to respond at a sixth-grade reading level improved readability while increasing the proportion of responses judged accurate and complete (85% vs. 77%) and reducing unacceptable responses (2% vs. 6%) (Hamzeh et al., 2026). Similarly, ChatGPT successfully simplified peer-reviewed ophthalmology articles from approximately a 15th- to a 7th-grade reading level, with the rewritten content considered appropriate by expert ophthalmologists (Kianian et al., 2024). The clinically actionable message, that readability is a prompt-engineering variable, not a fixed model property, is well supported. Figure 5 summarizes the gap between unprompted reading level and the recommended band, and what happened to content quality when the reading level was brought down by prompt instruction alone.

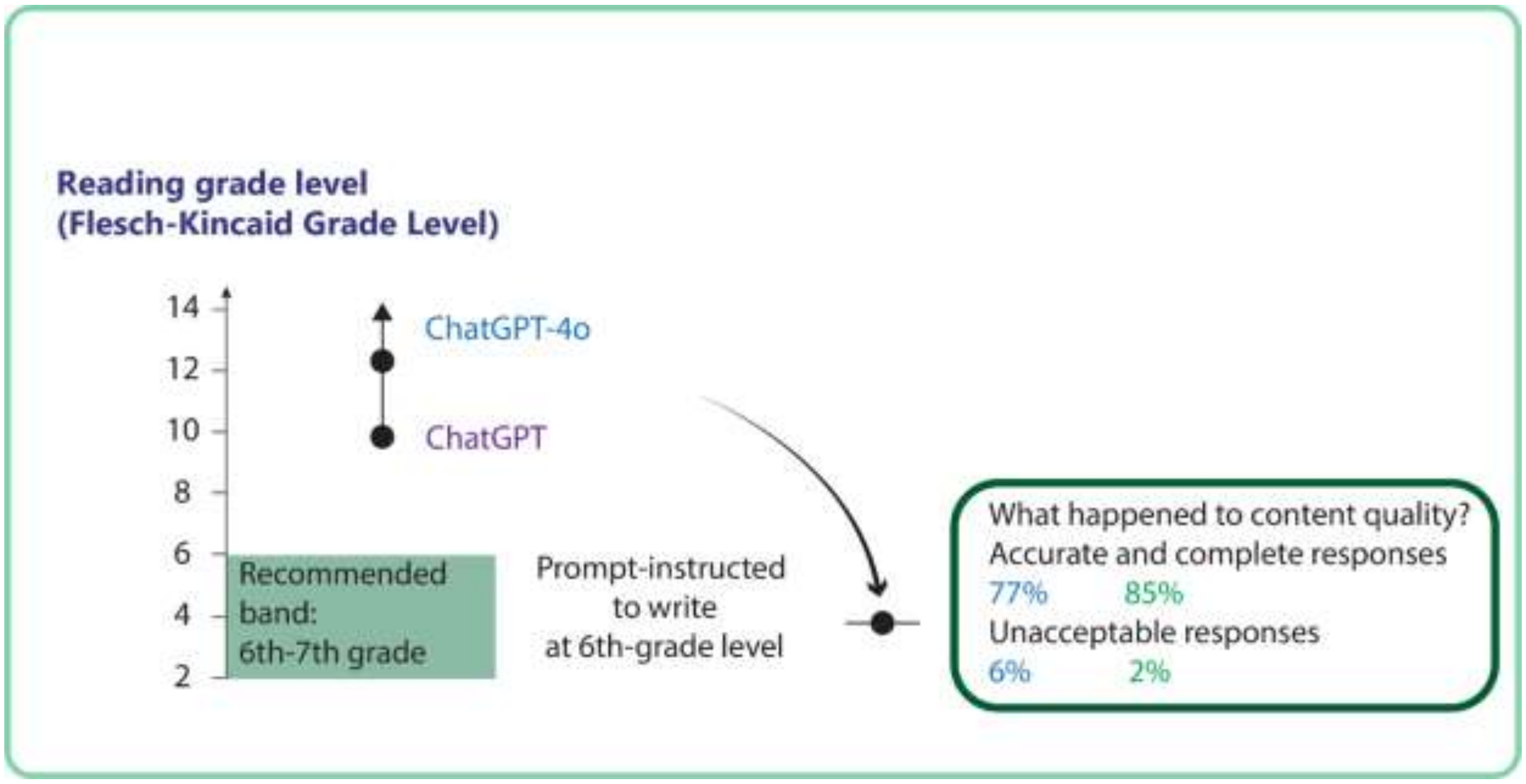

**Fig.** 5**.** Readability of large language model output behaves as a prompt variable rather than a fixed model property. Unprompted responses to common ophthalmic patient questions cluster near a tenth-grade reading level, and responses to real patient portal messages averaged a Flesch–Kincaid grade level of 12.1, both well above the sixth- to seventh-grade band recommended for patient education material. The Flesch–Kincaid grade level is a readability index expressing text difficulty as the United States school grade required to understand it. When the model was instructed within the prompt to write at a sixth-grade level, reading level fell into the recommended band while content quality was preserved: responses judged accurate and complete rose from 77 percent to 85 percent, and responses judged unacceptable fell from 6 percent to 2 percent. Data from Hamzeh et al. and Riazi Esfahani et al. (Hamzeh et al., 2026; Riazi Esfahani et al., 2025).

A major limitation of the current literature is that nearly all studies evaluate the quality of the generated response rather than its effect on patients. Whether LLM-generated education actually improves patient comprehension, reduces anxiety, increases treatment adherence, or influences health-seeking behavior remains largely unknown. As a result, current evidence demonstrates that LLMs can generate acceptable educational content, but provides little evidence that this translates into better patient outcomes(Bean et al., 2026).

**Appraisal note. The comparator problem.** A recurring limitation in the evaluation of LLM-generated responses is the use of reference materials that differ substantially from the responses being assessed, including institutional patient education leaflets, previously documented clinician responses, or answers originally developed for patients with different clinical histories and informational needs. Such comparisons may introduce systematic biases because the responses are produced under different conditions and serve distinct communicative purposes. Clinicians typically formulate advice within a specific clinical encounter, informed by the patient's medical history, individual circumstances, and contextual cues, whereas LLMs often generate responses to isolated questions without access to equivalent patient-specific information. Consequently, observed differences in response quality may partly reflect disparities in the information available, the intended audience, and the purpose of the communication, rather than differences in the underlying clinical capabilities of the two sources.

A further concern relates to the effectiveness of blinding in comparative assessments. Differences in writing style, response length, organization, and linguistic patterns may allow evaluators to infer whether a response was generated by an LLM or written by a clinician, even when the source is not explicitly disclosed. Such cues may influence judgments independently of clinical accuracy, completeness, or relevance. Therefore, nominal blinding does not necessarily ensure that assessments are free from source-related bias. When reference materials are not sufficiently comparable and evaluators can potentially identify the origin of responses, the validity of between-source comparisons may be compromised. These methodological considerations highlight the importance of contextually appropriate reference standards, comparable assessment conditions, and rigorous blinding procedures when evaluating the clinical utility of LLM-generated responses.

### 3.3 Diagnosis from clinical vignettes and case reports

Diagnostic evaluation accounts for roughly 15% of the ophthalmic corpus (Zhang et al., 2025) and has been performed across different diseases such as glaucoma, corneal disease, neuro-ophthalmology, uveitis, and rare eye diseases (Delsoz et al., 2024; Delsoz et al., 2023; Hu et al., 2023; Madadi et al., 2024; Rojas-Carabali et al., 2024).

Reported accuracies vary enormously, and the pooled estimate of 0.594 with $I^2$ = 94.5% is best read as a statement that these studies are not measuring the same thing. Subgroup pooling by subspecialty ranged from 0.483 (retina and vitreous) to 0.731 (external disease and cornea), with confidence intervals wide enough to overlap almost entirely (Zhang et al., 2025).

**Appraisal note. Vignettes are not patients.** Published case reports, the dominant source of diagnostic vignettes, have been written retrospectively by an author who already knows the diagnosis. The narrative is therefore pre-filtered for diagnostic relevance: irrelevant findings have been dropped, relevant findings foregrounded, and the differential implicitly narrowed. A model achieving 70% on such vignettes should not be interpreted as demonstrating 70% diagnostic accuracy in undifferentiated clinical presentations. Additionally, published cases are disproportionately unusual which cuts in the opposite direction. Collectively, these features introduce substantial spectrum bias, limiting the generalizability of reported diagnostic performance to routine clinical practice(Kanjee et al., 2023; Mulherin & Miller, 2002; Ransohoff

& Feinstein, 1978). These concerns are amplified by the limited use of real-world clinical data(Hager et al., 2024). Only about one-quarter of ophthalmic LLM studies used real patient data, and only 4 of 187 studies (2.1%) evaluated models within routine clinical practice rather than simulated scenarios (Zhang et al., 2025). Figure 6 traces what is discarded and what is emphasized as a real encounter is converted first into a published case report and then into a prompt.

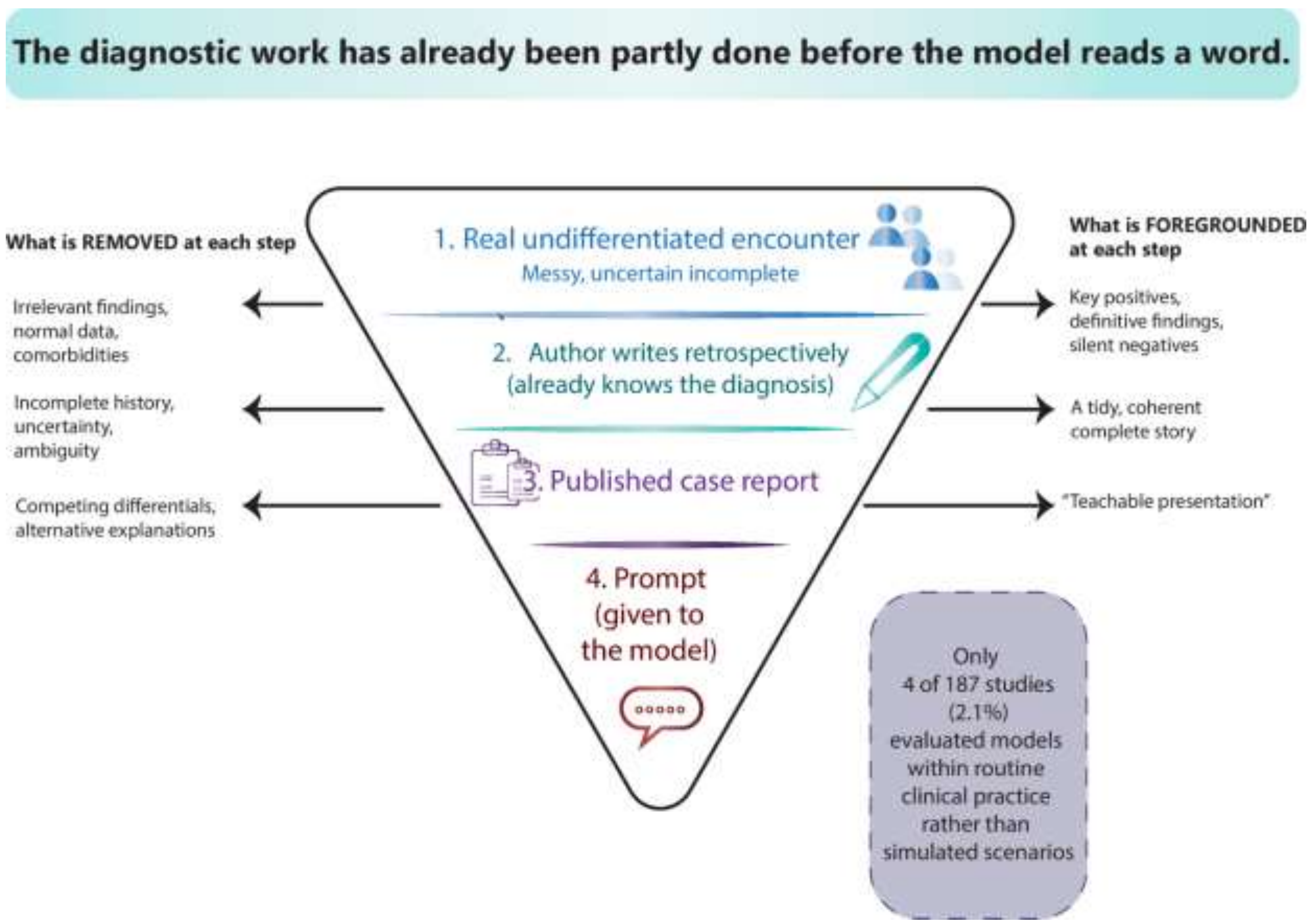

**Fig. 6.** How a clinical encounter becomes a prompt. At each step information is removed and other information is brought forward, so that part of the diagnostic work is already complete before the model receives any text. The real encounter is undifferentiated, incomplete and uncertain; the author writes retrospectively, already knowing the diagnosis; the published case report presents a coherent and teachable narrative; and the prompt preserves only that narrative. Performance on such items therefore measures reasoning over curated clinical narratives rather than diagnostic

performance in undifferentiated presentations. This is a form of spectrum bias, meaning that the mix of cases tested differs systematically from the mix encountered in the setting where the system would be used. Only 4 of 187 ophthalmic large language model evaluations (2.1 percent) assessed models within routine clinical practice rather than simulated scenarios. Data from Zhang et al. (Zhang et al., 2025).

### 3.4 Triage and urgency assessment

Patient triaging accounted for 1.6% of studies in the Zhang corpus (Zhang et al., 2025). This is where LLMs would deliver most value in resource-constrained systems, and where the evidence base is thinnest.

Early work compared GPT-4 and Bard on common ophthalmic complaints and assessed chatbot triage of ophthalmic conditions against clinician judgement (Lyons et al., 2024; Zandi et al., 2024). More recent studies are considerably stronger. Cardoso-Teixeira et al. evaluated five model systems on 3,831 real anonymized Portuguese-language referral vignettes across three independent runs, reporting baseline classification accuracy of 68.7% rising to 73.4% after supervised in-context learning, with peak accuracy of 79.5% and, importantly, a large gain in consistency for one model (66.8% to 93.8%). Performance exceeded 90% for common referral categories and degraded on rarer ones (Cardoso-Teixeira et al., 2026).

In another vignette-based comparative study, ChatGPT-5, Gemini 2.5 Pro, and Claude Opus 4.1 demonstrated comparable performance across urgency triage, and red flag recognition of 40 ophthalmic emergency scenarios, with no significant differences between models (Mittal & Aggarwal, 2026).

Two design features of the Cardoso-Teixeira study deserve emulation and are otherwise almost unseen in this literature: real-world clinical text rather than synthetic vignettes, and repeated independent runs with consistency reported as a primary outcome. Consistency is arguably more important than accuracy for triage deployment. A model that is right 80% of the time but unpredictably so is more dangerous in a referral pathway than one that is right 75% of the time reproducibly, because the latter's failure modes can be characterized and guarded.

**Appraisal note. Most studies treat all triage errors as equally important, even though they are not.** Missing a sight-threatening condition such as acute angle-closure glaucoma, giant cell arteritis, retinal detachment, or endophthalmitis can lead to permanent vision loss, whereas unnecessarily referring a patient with a benign condition such as a chalazion mainly results in an extra clinic visit. Despite this, most studies report only overall accuracy, which gives the same weight to both types of errors. Future studies should report sensitivity for sight-threatening conditions separately and, where possible, use evaluation metrics that assign greater weight to clinically serious errors. Equally important, future studies should evaluate whether LLM-assisted triage shortens the time to appropriate specialist care. Ultimately, reducing delays in treatment may be more clinically meaningful than small improvements in classification accuracy alone. We found no ophthalmic triage study that has adopted this approach. Figure 7 illustrates both points: why consistency across repeated runs is a safety outcome in its own right, and why a single overall accuracy figure cannot represent the clinical cost of a triage error.

A related and essentially untested dimension is whether triage or referral recommendations vary systematically by patient demographic characteristics rather than by clinical presentation. Work examining bias in AI chatbot ophthalmologist recommendations found both inaccuracy and bias in referral suggestions(Oca et al., 2023). Systematic differences in recommended screening intervals or referral urgency by patient demographic characteristics would be a serious harm and have not been systematically tested in ophthalmology.

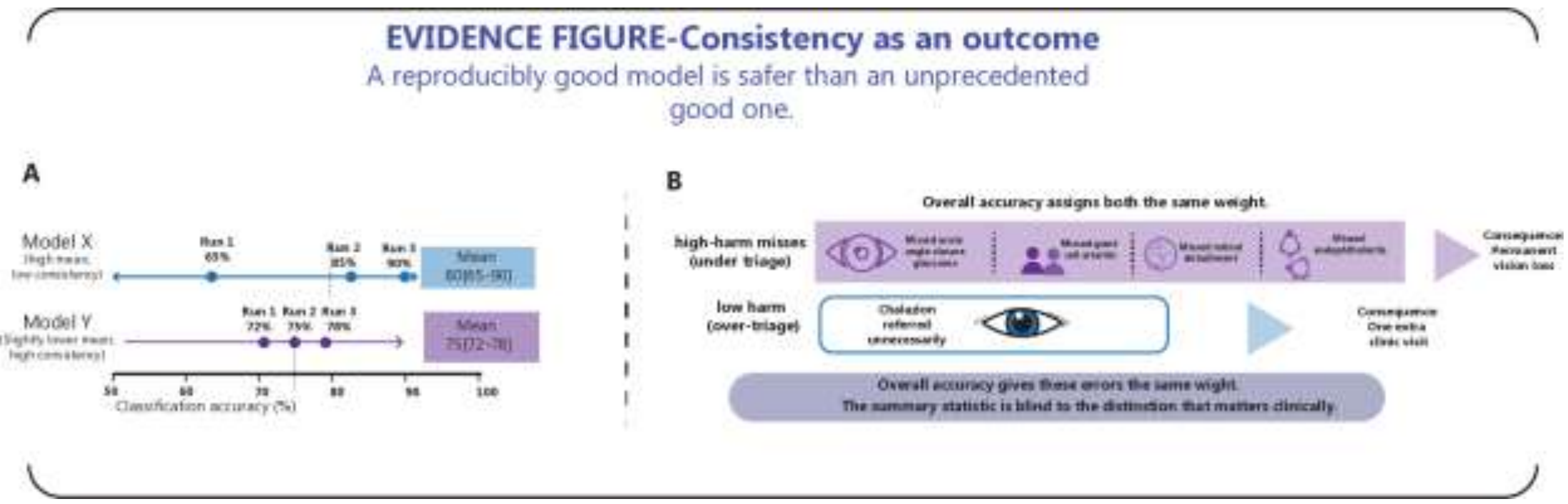

**Fig. 7.** Two triage outcomes that matter clinically and are rarely reported. (A) Consistency across repeated runs. Model X has the higher mean classification accuracy (80 percent) but varies widely between runs (65 to 90 percent), whereas Model Y has a slightly lower mean (75 percent) within a narrow range (72 to 78 percent). A reproducibly good model is safer within a referral pathway than an unpredictably better one, because its failure modes can be characterized and guarded against. Values are illustrative. (B) Asymmetry of triage errors. Failing to identify a sight-threatening condition such as acute angle-closure glaucoma, giant cell arteritis, retinal detachment or endophthalmitis may result in permanent vision loss, whereas referring a benign condition such as a chalazion unnecessarily results in one additional clinic visit. Overall accuracy assigns both errors the same weight and is therefore blind to the distinction that matters clinically.

### 3.5 Documentation and research support

Documentation and research support represent relatively understudied but practically important applications of LLMs in ophthalmology. Unlike diagnostic or triage tasks, these applications typically involve clinician oversight, potentially reducing the risks associated with erroneous outputs (Hua et al., 2023; Patel & Lam, 2023; Singh et al., 2023; Van Veen et al., 2024; Waisberg et al., 2023).

The documentation use case is plausibly the nearest-term clinical value, precisely because it is the lowest-risk: a clinician reviews and signs the output, the error-detection burden falls on someone qualified to bear it, although automation bias can weaken this safeguard, even among experienced readers(Dratsch et al., 2023; Goddard et al., 2012), and the counterfactual is a rushed human-written note. Despite its potential for near-term implementation, documentation support has received relatively little attention compared with diagnostic benchmarking.

Unlike documentation support, the literature on research assistance consistently identifies citation and content reliability as the major limitation. Across ophthalmology-specific and general medical evaluations, fabricated references, inaccurate bibliographic details, and hallucinations remain common even in current-generation models(Alkaissi & McFarlane, 2023; Gravel et al., 2023). Hua et al. evaluated ophthalmic scientific abstracts and references generated by AI chatbots and documented reference fabrication (Hua et al., 2023). This is consistent with the broader literature: analysis of 636 citations across 84 GPT-3.5 and GPT-4 generated documents found substantial rates of both outright fabrication and citation error in non-fabricated references (Walters & Wilder, 2023). An influential study also catalogued analyses reporting fabrication rates as high as 47% in generated medical references (Bhattacharyya et al., 2023). Comparative work on systematic review support found hallucination rates of 39.6%, 28.6% and 91.4% for GPT-3.5, GPT-4 and Bard respectively (Chelli et al., 2024).

Accordingly, LLM-generated references require manual verification against the original source, ideally using DOI-level validation, and LLMs should not be relied upon as the sole method of literature identification. Figure 8 contrasts the two applications: one in which a qualified reviewer stands between the output and the patient, and one in which nothing does.

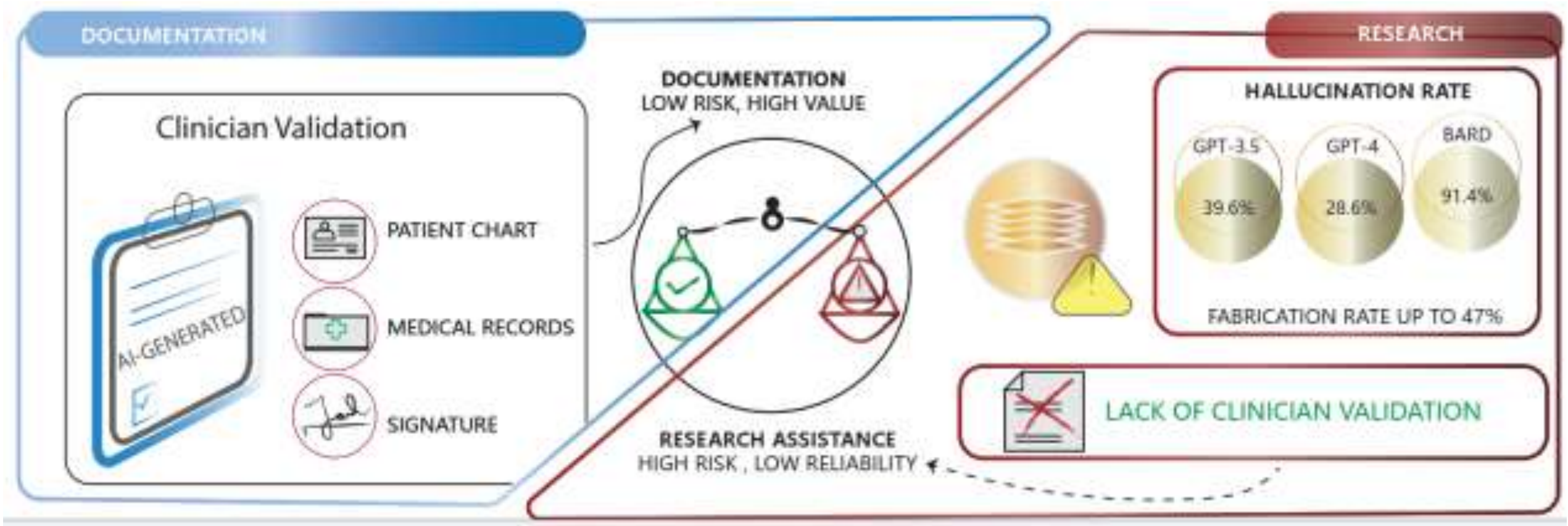

**Fig. 8.** Two non-clinical applications with opposite risk profiles. Documentation support is comparatively low risk and high value, because the output is reviewed and signed by a clinician and the burden of detecting error therefore falls on someone qualified to carry it. Research assistance is high risk and low reliability: usually no clinician validation step is built into the workflow, and the fluency that makes the output persuasive is the same property that makes fabricated material difficult to detect. Reported hallucination rates for reference generation were 39.6 percent for GPT-3.5, 28.6 percent for GPT-4 and 91.4 percent for Bard, with fabrication rates as high as 47 percent reported in generated medical references. Hallucination here denotes fluent and confident output that is fabricated or unsupported by any source. Data from Bhattacharyya et al. and Chelli et al. (Bhattacharyya et al., 2023; Chelli et al., 2024).

### 3.6 Multimodal image interpretation

Ophthalmology is an imaging specialty, yet this area has received surprisingly little research attention: only 10.2% of ophthalmic LLM studies involved images at all (Zhang et al., 2025).

Where multimodal capability has been tested, results are sobering. Xu et al. benchmarked GPT-4V on ophthalmic multimodal image analysis and titled the resulting paper around its clinical incapabilities (Xu et al., 2024). Mihalache et al. assessed chatbot diagnosis of clinical ophthalmic images with consistently modest results(Andrew Mihalache et al., 2024; A. Mihalache et al., 2024). Pooled image-text diagnostic accuracy in the Zhang meta-analysis was 0.407 versus 0.619 for text-only tasks(Zhang et al., 2025).

Another quantitative evaluation of four LLMs for diabetic retinopathy detection and grading from ultra-widefield fundus images, against Early Treatment Diabetic Retinopathy Study (ETDRS) ground truth established by three retina specialists, found performance insufficient for clinical use across four prompting strategies (Most et al., 2026).

It is important to separate two distinct technical lineages here, which the literature frequently conflates:

- **General-purpose multimodal LLMs** (GPT-4V/4o, Gemini, Claude) have been tested on ophthalmic images in a zero-shot setting, meaning they were used without ophthalmology-specific training. Their performance has generally been modest, which is not surprising because ophthalmic imaging is a highly specialized field that relies on subtle image details and is likely underrepresented in the large image-text datasets used to train these models.
- **Domain-specific vision-language and foundation models** RETFound, EyeCLIP, generalist ophthalmic vision foundation models, FFA-GPT and ChatFFA for fluorescein angiography interpretation, OphGLM, EYE-Llama, and integrated image-plus-language pipelines for primary diabetes care (Chen, Xu, et al., 2024; Chen, Zhang, et al., 2024; Z. Deng et al., 2024; Haghighi et al., 2025; Li et al., 2024; J. Qiu et al., 2024; Shi et al., 2025;

Zhou et al., 2023). These perform far better and represent the more plausible translational path (Sevgi et al., 2025). Unlike general-purpose multimodal LLMs, these models are pretrained or fine-tuned on large ophthalmic imaging datasets, allowing them to learn disease-specific visual representations.

Reviews that report "LLMs perform poorly on ophthalmic images" without this distinction are conflating a question about general-purpose chatbots with a question about purpose-built systems. The honest summary is: general-purpose multimodal models are not currently adequate for ophthalmic image interpretation; domain-specific vision-language models show substantially more promise but have themselves been evaluated almost entirely retrospectively on curated datasets.

**Appraisal note. Limited Evidence for Open-Source LLM Deployment in Ophthalmology.** Open-source models have been evaluated in only 13.4% of ophthalmic LLM studies overall, although they were used in 77.3% of studies that developed or fine-tuned a model rather than only evaluating existing systems (Zhang et al., 2025). This limited attention is important because open-weight models may be the most practical option for many healthcare systems, particularly those with limited resources or strict data privacy requirements. These models can be deployed locally within hospitals, reducing concerns related to data sharing and external cloud services(Riedemann et al., 2024; Wiest et al., 2024). However, evidence regarding their performance and clinical applicability in ophthalmology remains limited, representing a clear opportunity for future research.

### 3.7 Uncertainty in ophthalmic LLMs

**Why uncertainty matters.** Every finding in Sections 3.1 through 3.6 describes what these models get right on average. Few studies, however, assess whether a model can reliably distinguish cases in which its predictions are likely to be correct from those in which its outputs should be interpreted with caution. This represents a distinct dimension of model performance that cannot be adequately captured by accuracy alone (Omar et al., 2025; Van Calster et al., 2019). A system that fails silently, at the moment it is least confident, converts a knowledge gap into a missed diagnosis. Calibration (the correspondence between a model's stated or implicit confidence and its actual correctness) was assessed in 2.7% of the 187 studies in the Zhang corpus, against accuracy reporting in 94.1% (Zhang et al., 2025). The field has, in effect, measured what these systems know and almost never measured what they know they do not know.

**What uncertainty-aware LLMs should actually do.** An uncertainty-aware LLM should do two things that current LLMs usually do not.

*First, it should distinguish between two different types of uncertainty.* Aleatoric uncertainty comes from the case itself. For example, an image may be blurry, a lesion may have an unclear border, or even experts may disagree on the diagnosis. In these situations, uncertainty is expected because the available information is inherently limited. In contrast, epistemic uncertainty reflects the model's own lack of knowledge. This usually occurs when the model encounters rare diseases, unusual clinical presentations, or data that differ from what it has seen during training. Unlike aleatoric uncertainty, epistemic uncertainty can often be reduced by improving the model or providing more representative training data(Hüllermeier & Waegeman, 2021; Leibig et al., 2017).

These two types of uncertainty have different meanings. A model that is uncertain about an ambiguous fundus image may be behaving appropriately. However, if it is equally uncertain about a classic case of proliferative diabetic retinopathy, this suggests a limitation of the model rather than a difficult case. Most reported only overall accuracy, making it impossible to determine whether the observed errors were caused by genuinely problematic, incomplete input or by gaps in the model's knowledge.

*Second, an uncertainty-aware LLM should know when not to answer confidently.* When confidence is low, the safest response is not to guess but to acknowledge the uncertainty and recommend human review or additional clinical evaluation(Dvijotham et al., 2023; Kompa et al., 2021). For screening and triage applications, this should be considered a core safety feature rather than a limitation(Griot et al., 2025).

**Why current confidence measures fail.** The confidence signals used in almost all ophthalmic LLM studies are the ones that the broader uncertainty-quantification literature considers least reliable(Gao et al., 2025; Savage et al., 2025; Xiong et al., 2024). One common approach is to use the probability assigned to the generated words (tokens) as a measure of confidence. For example, repeated responses alternating between proliferative diabetic retinopathy, retinal vein occlusion, and ocular ischemic syndrome indicate genuine diagnostic uncertainty because the model cannot consistently settle on the same diagnosis. Token-level confidence does not measure this. Instead, it measures the probability of individual generated words. Consequently, two responses expressing exactly the same diagnosis (for example, "proliferative diabetic retinopathy" and "the findings are most consistent with proliferative diabetic retinopathy") may receive different confidence estimates simply because they use different wording(Kuhn et al., 2023).

This helps explain a phenomenon already encountered throughout this review under different names. What the uncertainty-quantification literature refers to as medical hallucination (fluent, confident, but fabricated or unsupported output) is not a separate failure mode from the incorrect reasoning and fabricated references discussed earlier. Rather, these problems share the same underlying cause: LLMs are trained to generate the most plausible continuation of text, not to determine whether the information is true(Kalai et al., 2025; Kuhn et al., 2023). Consequently, a fabricated citation may look just as convincing as a genuine one, and an incorrect clinical explanation may be presented with the same confident language as a correct one. Recent work has shown that meaning-based uncertainty measures, such as semantic entropy, are substantially better at detecting these hallucinations than conventional token-level confidence because they measure uncertainty in the underlying meaning rather than in the exact wording of the response (Farquhar et al., 2024).

The same problem applies to the model's stated reasoning. Asking an LLM to explain its diagnosis produces fluent justification, but there is little basis for treating that justification as a faithful account of how the answer was actually produced(Turpin et al., 2023). The study discussed in Section 3.1 is the clearest available evidence: GPT-4 selected the correct multiple-choice answer in 82.5% of items, yet expert raters identified incorrect reasoning in 42% of the free-text justifications the same model produced (Chen et al., 2026). This matters because a confident, fluent, but unfaithful explanation is more dangerous than no explanation at all: an opaque model invites appropriate caution, whereas a plausible-looking rationale invites trust it has not earned(Ghassemi et al., 2021).

**Possible solutions.** Semantic entropy directly addresses the paraphrasing problem above and has not, to our knowledge, been applied in any ophthalmic evaluation.

Conformal prediction may offer a practical and promising approach to addressing this limitation, although its application in the ophthalmic LLM literature appears to remain unexplored or insufficiently documented (Mohri & Hashimoto, 2024; Quach et al., 2024). Instead of one answer with an unreliable confidence score, it returns a prediction set with a formal statistical coverage guarantee: for a diagnostic task, a ranked shortlist of differentials designed to contain the correct diagnosis with a predefined statistical coverage (e.g. 95%) under appropriate assumptions. This maps with unusual precision onto the diagnostic-vignette and triage tasks of Sections 3.3 and 3.4, where the clinically useful output may not be the model's single best guess but the set of possibilities that cannot yet be safely excluded.

**Proposed response framework for future ophthalmic LLMs.** Rather than returning a diagnosis alone, future ophthalmic LLMs should provide outputs that explicitly communicate uncertainty and reliability.

Figure 9 contrasts current practice with an uncertainty-aware alternative.

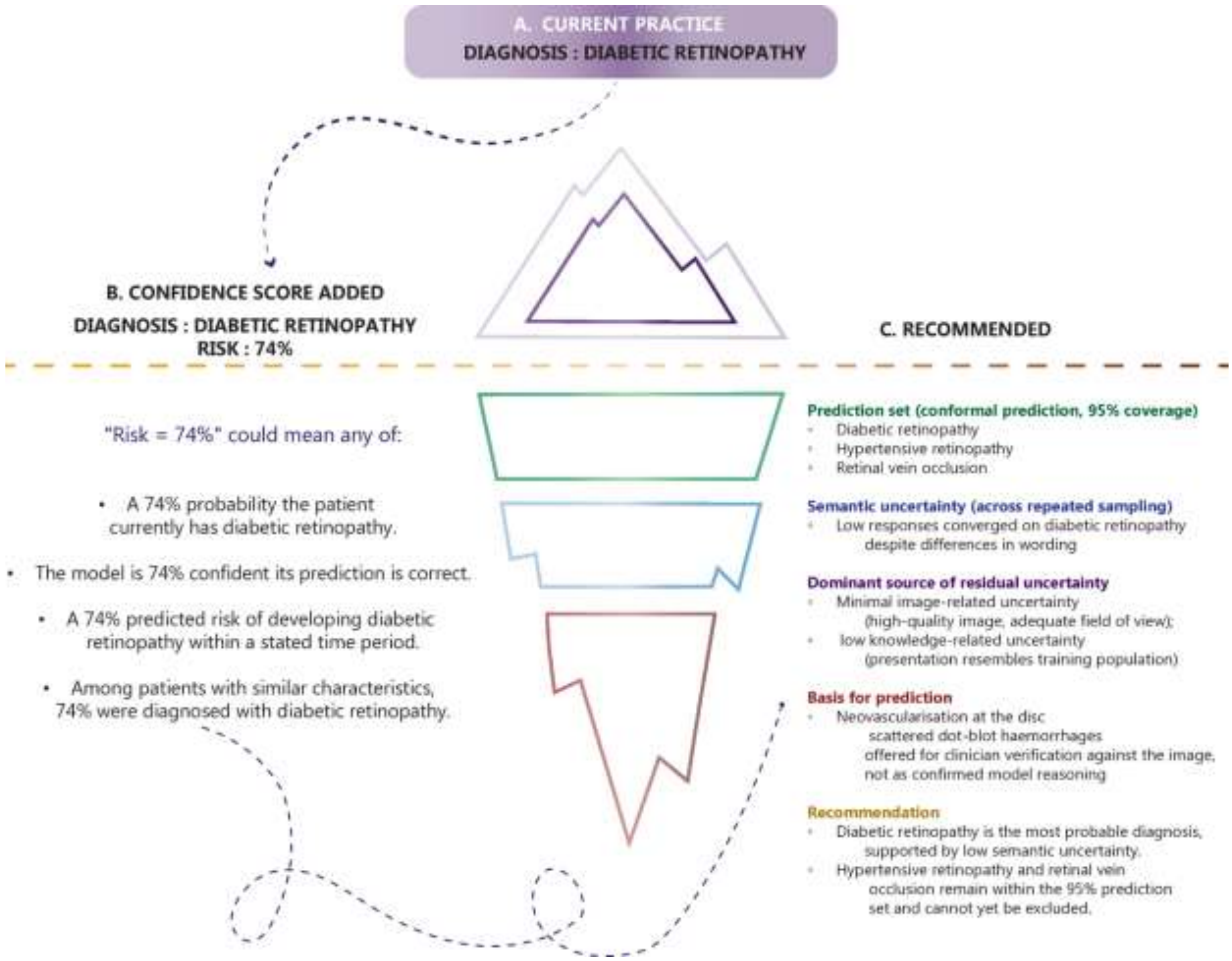

**Fig 9.** Progression from current practice (A) to an uncertainty-aware clinical output (C). A bare confidence percentage (B) is ambiguous between at least four distinct clinical meanings. The recommended format (C) communicates what the model concludes, how reliable that conclusion is, why residual uncertainty exists, and what further evaluation would reduce it.

This framework changes the role of the LLM from an answer generator into a decision-support system. Rather than presenting a single diagnosis with undefined confidence, it communicates what the model believes, how reliable that belief is, why uncertainty exists, and what additional evaluation is required before uncertainty can be reduced. Such outputs are considerably more aligned with real clinical reasoning than current confidence scores or single-label predictions.

## 4. Methodological appraisal: the SIGHT framework

We constructed SIGHT inductively; from failure modes observed across the corpus reviewed in Section 3. It comprises five domains and twenty items: **S**ample and item provenance, **I**ndependence and blinding, **G**round truth and comparator fairness, **H**andling of paired and repeated-measures data, and **T**ransparency and trustworthiness.

**Scoring.** Each item is scored Yes = 2, Partial = 1, No or Not reported (NR) = 0. Non-reporting is scored identically to non-compliance: a design feature that cannot be verified from the manuscript cannot be credited. Items deemed not applicable (NA) to a given study design are excluded from both the numerator and denominator. For each domain, the percentage score is calculated as the points earned divided by twice the number of applicable items, multiplied by 100. The overall SIGHT score is calculated as the unweighted mean of the domain percentage scores. The number of applicable items should be reported for each domain. The maximum applicable score is 40, comprising S = 10, I = 6, G = 6, H = 8, T = 10.

**Interpreting and applying the profile.** Eleven of the twenty items are NA-eligible, so applicable denominators vary by design. Two consequences follow, and both have practical remedies.

First, because domains contain unequal numbers of items, raw summation implicitly weights the larger domains. Scores should therefore be domain-normalized: each domain expressed as a percentage of its own applicable maximum, and the overall score computed as the mean of domains with at least one applicable item.

Second, domains are not exchangeable in what their failure implies. We designate **H1 (analysis matched to data structure)** and **T1 (model version identified)** as critical items: a score of 0 on either classifies a study as high concern irrespective of its total, since the first invalidates the study's inferential claim and the second precludes replication. Overall judgment is therefore categorical; low concern (≥70% domain-normalized, no critical item at 0, no domain below 40%), moderate concern (40–70%, or any domain below 40%), high concern (<40%, or any critical item at 0).

### 4.1 S: Sample and item provenance

**S1. Are test items drawn from real clinical encounters, or synthesized?**

Synthetic and published-case items are not inherently inferior, but they measure a different construct: performance on curated clinical narratives in which information has already been selected, organized, and edited by someone who knows the diagnosis. Real encounters contain incomplete histories, irrelevant findings, and evolving presentations.

- **Yes (2):** items derive from real clinical encounters.
- **Partial (1):** items are synthetic, published-case-derived, or investigator-written, and the study explicitly limits its conclusions to curated-case performance rather than clinical performance.
- **No / NR (0):** synthetic or published-case items with conclusions generalized to clinical practice without qualification; or the source is described too vaguely to determine.

**S2. Has training-data contamination been handled?**

Any study using a commercial question bank, published case series, or public patient-education website must address whether those items sit in the model's pre-training corpus. Contamination inflates apparent accuracy without any corresponding gain in reasoning ability.

- **Yes (2):** contamination is either prevented or quantified. Prevented: items were created after the model's training cutoff, or are demonstrably inaccessible to the web. Quantified: The study reports whether evaluation items predate the model's training cutoff and describes how training-data contamination was minimized.
- **Partial (1):** contamination is explicitly acknowledged as a limitation, but no mitigation or sensitivity analysis is undertaken.
- **No / NR (0):** publicly available items are used with no mention of contamination.
- **NA:** the task supplies all necessary information in-context and evaluates reasoning over the provided material rather than recall from training data (e.g., summarizing a provided operative note or simplifying a supplied patient information leaflet).

**S3. Is the sample size justified, with the unit of analysis specified?**

The unit of analysis is frequently ambiguous (powered on items, responses, raters, or runs) and that ambiguity propagates into the statistical analysis (see 4.4).

- **Yes (2):** an a priori power or precision calculation is reported, with the unit of analysis explicitly named.
- **Partial (1):** the unit of analysis is clearly identifiable and the sample size is justified by reference to feasibility, precedent, or expected precision, without a formal calculation.

- **No / NR (0):** no justification offered, or the unit of analysis cannot be determined (e.g. "30 questions were assessed by four raters across three runs" with no statement of what n refers to in the analysis).

**S4. Does the item set represent the spectrum of cases the system would encounter in the intended deployment setting?**

Published case reports, teaching collections, and question banks are each enriched for diagnostically informative, unusual, or pedagogically clean material, and so do not reflect the prevalence, complexity, or case mix of routine practice. The requirement is not that items be unselected, but that selection be described and its relationship to the deployment population addressed.

- **Yes (2):** items are drawn consecutively or unselectively from the target setting, or the distribution of diseases, case severity, or urgency in the evaluation set is shown to be representative of the intended deployment population.

- **Partial (1):** the item source is clearly described, and this is acknowledged as a limitation on generalizability.

- **No / NR (0):** an enriched or curated source is used and results are presented as representative of clinical performance without qualification; or the source is insufficiently described to judge.

**S5. Is the language of evaluation reported and appropriate for the intended deployment population?**

Performance varies substantially across languages and results in one language should not be assumed to transfer. Where intended users or patients are not monolingual in the evaluation language, single-language evaluation constrains what the study can claim.

- **Yes (2):** the evaluation language is stated and either performance is assessed across the languages relevant to the intended setting, or single-language evaluation is explicitly justified by reference to that setting.
- **Partial (1):** the language is stated, evaluation is single-language, and the limitation is acknowledged without justification tied to deployment context.
- **No / NR (0):** evaluation is single-language with conclusions generalized beyond it without qualification; or the evaluation language is not stated at all.

### 4.2 I: Independence and blinding

**I1. Were raters blinded to response source, with blinding integrity verified?**

LLM prose has recognizable stylistic markers (characteristic length, enumerated structure, hedging) that allow source identification without clinical judgement. Blinding therefore requires active mitigation, not merely withholding labels.

- **Yes (2):** raters were blinded and mitigation is described (normalizing formatting and length across arms, stripping enumeration and headers), and blinding success is verified, ideally by asking raters to guess the source and reporting their accuracy.

- **Partial (1):** raters were blinded to source labels, but no formatting normalization or verification is reported.
- **No / NR (0):** raters were unblinded, or blinding is not mentioned in a study whose design requires it.
- **NA:** no human rating: responses scored automatically against a fixed answer key, or against automated metrics only.

**I2. Were two or more independent raters used, with agreement quantified?**

- **Yes (2)**
- **Partial (1):** multiple raters used, with agreement reported as a bare point estimate or raw percentage agreement only, without CI.
- **No / NR (0):** single rater, or multiple raters with no agreement statistic reported.
- **NA:** no subjective rating involved (automated scoring against an objective key).

**I3. Is rater assignment free of confounding with the comparison of interest?**

In nested designs where each rater assesses only one model, rater leniency is perfectly confounded with model identity, and any observed between-model difference is uninterpretable.

- **Yes (2):** all raters assessed all arms (fully crossed design), or a nested design includes an anchoring mechanism (a common item set rated by every rater) permitting estimation and adjustment of rater-specific leniency.
- **Partial (1):** a nested or partially crossed design is used, the confounding is explicitly acknowledged, but no anchoring or statistical adjustment is applied.

- **No / NR (0):** a nested design is used with between-arm comparisons reported as though raters were interchangeable, with no acknowledgement; or the assignment structure cannot be determined.
- **NA:** no human raters, or only one arm evaluated with no between-arm comparison.

### 4.3 G: Ground truth and comparator fairness

#### G1. Is the reference standard defensible and its derivation reported?

- **Yes (2):** the reference standard's derivation is fully specified: for expert-panel ground truth, the number and subspecialty match of experts, their agreement, and the adjudication procedure; for record-derived truth, the criteria used and acknowledgement that the recorded diagnosis may itself be wrong; for answer keys, the source and its editorial provenance.
- **Partial (1):** the reference standard's source is stated but its derivation is incompletely specified (e.g. "confirmed by two retina specialists" with no agreement statistic or adjudication rule).
- **No / NR (0):** the reference standard is asserted without derivation.

#### G2. Is the human comparator concurrent and tested on the identical item set?

Comparison against historical benchmarks is not a controlled comparison: those figures reflect different examinee populations, testing conditions, time pressure, and stakes.

- **Yes (2):** human comparators were tested concurrently on the identical item set, under stated conditions.
- **Partial (1):** a human comparator is used but drawn from a non-concurrent or non-identical source (historical pass rates, vendor norms, previously published clinician answers), and this limitation is explicitly acknowledged.
- **No / NR (0):** human-comparative claims ("comparable to ophthalmologists," "exceeded the human pass rate") are made on the basis of non-concurrent benchmarks without qualification.
- **NA:** the study makes no human-comparative claim, evaluating model performance against a reference standard only.

*Applicability:* All domains, but note the NA condition carefully; a study that never claims human comparability should not be penalized for lacking a comparator. The distinction is between *not comparing* (NA) and *comparing badly* (No).

**G3. Did model and human arms receive equivalent clinical information?**

- **Yes (2):** both arms received identical inputs, or asymmetries are documented and either equalized or accounted for analytically.

- **Partial (1):** an information asymmetry exists (the clinician had the chart, examination, or prior relationship; the model had a text string) and is explicitly acknowledged as limiting interpretation.
- **No / NR (0):** an asymmetry exists and is not acknowledged, so the study measures the information difference while attributing it to reasoning capability.
- **NA:** no human comparator arm.

### 4.4 H: Handling of paired and repeated-measures data

Most comparative LLM studies evaluate multiple models on the same set of questions or clinical cases. Because each model is tested on identical items, the observations are paired rather than independent, and the analysis should reflect this design. Treating paired observations as independent ignores the correlation between model responses and can lead to incorrect estimates of variability, misleading p-values, inappropriate confidence intervals, and loss of statistical power. It may also exaggerate or obscure differences between models, making conclusions about comparative performance unreliable. Studies should therefore use statistical methods appropriate for paired or repeated-measures data, or explicitly justify why an independent analysis is appropriate.

#### H1. Is the statistical analysis matched to the paired or repeated-measures structure?

When multiple models answer the same items, the resulting accuracy vectors are repeated measures on shared items, not independent samples. Item difficulty is a shared unmeasured influence on all arms.

- **Yes (2):** analysis matches the data structure.
- **Partial (1):** a paired test is used where a mixed model would better separate item and rater effects, or the design is treated as paired in some analyses and independent in others without explanation.
- **No / NR (0):** independent-samples tests applied to paired data. Also scored 0 where the analysis performed cannot be determined.
- **NA:** a single model evaluated with no between-arm or between-condition comparison, and no repeated measurement.

**A further point on clustering.** Items drawn from the same subspecialty, the same question bank, the same source website or multiple runs from the same model are not independent of one another. Where the item set has such structure, it should be modelled as a random effect or via cluster-robust standard errors rather than ignored.

**H2. Is multiplicity controlled?**

- **Yes (2):** a primary outcome is pre-specified and multiplicity is formally controlled, with the number of comparisons stated.
- **Partial (1):** correction is applied to some comparison families and not others.

- **No / NR (0):** multiple pairwise comparisons across models and outcome dimensions are reported uncorrected.
- **NA:** a single pre-specified comparison only.

**H3. Are effect sizes and confidence intervals reported alongside *p*-values?**

- **Yes (2):** effect estimates with confidence intervals for all primary comparisons.
- **Partial (1):** effect sizes reported without confidence intervals, or CIs given for some comparisons only.
- **No / NR (0):** *p*-values reported alone. Given typical sample sizes in this literature, a bare significance statement conveys almost nothing about clinical magnitude.

**H4. Was potential demographic bias in model outputs evaluated?**

The concern is whether the model's responses change because of patient characteristics (such as age, sex, or ethnicity) rather than because of clinically relevant differences. For example, two patients with the same presentation should not receive different urgency recommendations or management advice simply because they differ in age, sex, or ethnicity, unless there is a clear clinical reason.

- **Yes (2):** The study evaluates whether model performance or responses differ according to at least one patient demographic characteristic. This may include comparing performance across demographic groups or using identical clinical vignettes that differ only in age, sex, or ethnicity.
- **Partial (1):** Patient demographic characteristics are reported but their relationship with model performance is not analyzed.

- **No / NR (0):** Patient demographic information is included in the evaluation cases or prompts but is neither reported nor analyzed.
- **NA:** The evaluation items contain no patient demographic information.

*Applicability by domain:*

- Examinations: Usually NA, as examination questions typically do not include patient demographic information.
- Patient education: Assess whether the quality, readability, or recommendations vary according to patient demographics.
- Diagnosis: Assess whether changing only demographic characteristics (while keeping the clinical presentation identical) changes the model's diagnosis.
- Triage: Assess whether urgency or referral recommendations differ according to patient demographics despite identical clinical presentations.
- Documentation: Assess whether the tone, level of detail, or completeness of generated documentation differs according to patient demographics.

### 4.5 T: Transparency and trustworthiness

#### T1. Is the exact model version, access date, and interface reported?

Providers ship revisions under unchanged public names; several distinct snapshots of a single model may be released within one year.

- **Yes (2):** the precise version string or snapshot identifier, the access or query date, and the interface used (web interface versus application programming interface [API]) are all reported.
- **Partial (1):** major version and approximate date given ("GPT-4, accessed March 2024") without snapshot identifier or interface.
- **No / NR (0):** only a model family named ("ChatGPT," "Gemini") with no version or date.

**T2. Is the stability of the performance estimate established?**

LLM output is stochastic. A study querying each item once has drawn a single observation from a distribution and reported it as a point estimate. Setting temperature to zero is a partial mitigation only, so determinism should be demonstrated rather than assumed.

- **Yes (2):** each item queried multiple times with both central tendency and dispersion across runs reported; generation parameters (temperature, top-p, seed) stated where accessible, or their inaccessibility declared for web-interface studies.
- **Partial (1):** multiple runs performed but only aggregate results reported without dispersion; or a single query per item with generation parameters fully disclosed and stochasticity explicitly acknowledged as a limitation.
- **No / NR (0):** single query per item with no parameter reporting and no acknowledgement of stochasticity; or determinism asserted on the basis of temperature = 0 alone.

**T3. Are prompts disclosed verbatim, with prompt sensitivity considered?**

- **Yes (2):** full prompts reproduced verbatim in text or supplement, and sensitivity to plausible prompt variants assessed or explicitly addressed.
- **Partial (1):** prompts disclosed verbatim with no sensitivity analysis.
- **No / NR (0):** prompts paraphrased, summarized, or omitted, so the study cannot be replicated.

**T4. Where the study claims clinical usefulness, is model confidence evaluated?**

Accuracy shows how often a model is correct. Calibration shows whether the model is more confident when it is correct and less confident when it is wrong. This item applies only if the study makes claims about clinical use, clinical readiness, or suitability for diagnosis, triage, or screening. Studies that report benchmark performance without making these claims are not expected to assess calibration. For example, if an LLM labels 100 fundus photographs as diabetic retinopathy with 90% confidence, a well-calibrated model should be correct in about 90 of those cases. If only 60 are truly diabetic retinopathy, the model is overconfident despite appearing highly certain.

- **Yes (2):** The study evaluates the relationship between model confidence and correctness. For triage or screening applications, it also reports how the model handles uncertain cases, such as abstaining from answering or recommending clinician review.
- **Partial (1):** The study reports model confidence but does not compare it with correctness. Alternatively, uncertainty or abstention is described but not quantified, consistency across repeated runs is used instead of a direct calibration assessment, or the authors acknowledge that the lack of calibration limits their conclusions about clinical use.

- **No / NR (0):** The study claims the model is suitable for clinical use, or ready for deployment without evaluating model confidence or acknowledging this limitation.
- **NA:** The study limits its conclusions to the measured task (e.g., benchmark performance) and makes no claims about clinical use, safety, or deployment.

**T5. Where the study draws conclusions from the model's explanations, are those explanations evaluated?**

A correct answer does not necessarily mean the model used correct reasoning. Likewise, a convincing explanation or attention map does not prove that the model actually based its decision on those reasons. Studies that claim a model is interpretable, transparent, or clinically trustworthy because of its explanations should evaluate those explanations directly, rather than assume they are reliable.

This item applies only when the study uses the model's explanations to support its conclusions. Studies that evaluate only the final answer, or provide example explanations without drawing conclusions from them, are not expected to meet this criterion.

- **Yes (2):** The study evaluates the quality or validity of the model's explanations independently of whether the final answer is correct. Examples include expert assessment of the reasoning, testing whether changing the explanation changes the prediction, or comparing highlighted image regions with expert-annotated diagnostic features.
- **Partial (1):** The study rates explanations as plausible or useful but does not evaluate whether they reflect the model's actual reasoning. Alternatively, the authors acknowledge that the lack of such evaluation limits their conclusions.

- **No / NR (0):** The study claims the model is interpretable, transparent, or clinically trustworthy based on its explanations or attention maps, but provides no evaluation of whether those explanations are valid.
- **NA:** The study evaluates only the model's final outputs and does not base any conclusions on its explanations.

A summary of the SIGHT framework is provided in Table1.

**Table 1.** The SIGHT framework.

| Domain (max) | Item | What it asks | Not applicable when |
|---|---|---|---|
| **S:** Sample and item provenance (10) | **S1** | Are items drawn from real clinical encounters, or synthesized? | - |
| | **S2** | Has training-data contamination been handled? | Task supplies all information in-context and tests reasoning over provided material, not recall |
| | **S3** | Is sample size justified, with the unit of analysis specified? | - |
| | **S4** | Does the item set represent the spectrum of cases in the intended deployment setting? | - |
| | **S5** | Is the evaluation language reported and appropriate for the deployment population? | - |
| **I:** Independence and blinding (6) | **I1** | Were raters blinded to response source, with blinding integrity verified? | No human rating; automated scoring only |
| | **I2** | Were ≥2 independent raters used, with agreement quantified? | No subjective rating involved |
| | **I3** | Is rater assignment free of confounding with the comparison of interest? | No human raters, or single arm with no between-arm comparison |
| **G:** Ground truth and comparator fairness (6) | **G1** | Is the reference standard defensible and its derivation reported? | - |
| | **G2** | Is the human comparator concurrent and tested on the identical item set? | Study makes no human-comparative claim |
| | **G3** | Did model and human arms receive equivalent clinical information? | No human comparator arm |
| **H:** Handling of paired and repeated-measures data (8) | **H1** | Is the analysis matched to the paired or repeated-measures structure? | Single model, no between-arm comparison, no repeated measurement |
| | **H2** | Is multiplicity controlled? | Single pre-specified comparison only |
| | **H3** | Are effect sizes and confidence intervals reported alongside p-values? | - |
| | **H4** | Was potential demographic bias in model outputs evaluated? | Items contain no patient demographic information |
| **T:** Transparency and trustworthiness (10) | **T1** | Is the exact model version, access date, and interface reported? | - |
| | **T2** | Is the stability of the performance estimate established across repeated queries? | - |
| | **T3** | Are prompts disclosed verbatim, with prompt sensitivity considered? | - |
| | **T4** | Where clinical usefulness is claimed, is model confidence evaluated against correctness? | Conclusions confined to the measured task; no clinical-use, safety, or deployment claim |
| | **T5** | Where conclusions rest on the model's explanations, are those explanations evaluated? | Only final outputs evaluated; no conclusions drawn from explanations |

**5. Limitations**

This is a narrative review and is therefore susceptible to selection bias in the studies discussed. We did not perform duplicate screening, did not register a protocol, and did not attempt exhaustive capture; readers requiring a complete enumeration should consult the existing systematic reviews.

The SIGHT framework should be regarded as a preliminary methodological framework rather than a validated risk-of-bias instrument. It was derived inductively from observed failure modes in the reviewed corpus, not through formal Delphi consensus, and its items have not been tested for inter-rater reliability(Diamond et al., 2014; Moher et al., 2010). We present it as a structured argument and a practical checklist, not as a validated instrument, and we would welcome formal development.

Finally, this review shares the currency problem it criticizes. Our search cutoff is stated, and model capabilities will have changed by publication. We have tried to mitigate this by emphasizing design principles, which are stable, over performance figures, which are not.

## 6. Conclusions

Large language models have reached a level of performance that makes questions of evaluation more important than questions of capability. Across ophthalmology, benchmark accuracy has improved rapidly, but evaluation methodology has not kept pace. Common weaknesses (including reliance on curated datasets, inadequate handling of paired study designs, limited assessment of uncertainty and calibration, insufficient attention to comparator fairness, and frequent overinterpretation of benchmark results) limit confidence in claims of clinical readiness.

The SIGHT framework was developed to address these recurring methodological shortcomings. Rather than introducing another reporting checklist, it provides a structured appraisal framework focused on whether the conclusions of an LLM evaluation are justified by its design, analysis, and interpretation. By emphasizing methodological validity alongside reporting quality, SIGHT complements existing guidance such as TRIPOD-LLM (Transparent Reporting of a multivariable prediction model for Individual Prognosis Or Diagnosis–Large Language Models), CHART (Chatbot Assessment Reporting Tool), and AQAT:RoB.

We do not present SIGHT as a validated risk-of-bias instrument. Instead, it is a framework based on recurring methodological limitations identified across the current ophthalmic LLM literature. Future work should refine its content through expert consensus, evaluate its inter-rater reliability, and examine its performance across broader medical specialties. As LLMs continue to evolve, improving how they are evaluated may be more important than incremental improvements in model accuracy. A trustworthy evaluation framework is essential if these systems are to be translated safely from benchmark studies into routine ophthalmic care.

**List of Abbreviations**

AI: Artificial Intelligence

API: Application Programming Interface

AQAT:RoB: Alberta Quality Assessment Tool: Risk of Bias

BCSC: Basic and Clinical Science Course

CHART: Chatbot Assessment Reporting Tool

CI: Confidence Interval

EQUATOR: Enhancing the QUAlity and Transparency Of health Research

ETDRS: Early Treatment Diabetic Retinopathy Study

FFA: Fundus Fluorescein Angiography

GPT: Generative Pre-trained Transformer

LLM: Large Language Model

NA: Not Applicable

NR: Not Reported

OKAP: Ophthalmic Knowledge Assessment Program

PRISMA: Preferred Reporting Items for Systematic Reviews and Meta-Analyses

SANRA: Scale for the Assessment of Narrative Review Articles

SIGHT: Sample and item provenance, Independence and blinding, Ground truth and comparator fairness, Handling of paired and repeated-measures data, and Transparency and trustworthiness

TRIPOD-LLM: Transparent Reporting of a multivariable prediction model for Individual Prognosis Or Diagnosis–Large Language Models

**Acknowledgments**

The authors used Claude and Grammarly in order to facilitate paraphrasing and grammar checking, during the preparation of this work. After using this tool, the authors reviewed and edited the content as needed and take full responsibility for the content of the publication.

**Declarations**

**Competing interests**

The authors declare that there is no conflict of interest.

**Funding Sources**

None.

**Authors' contributions**

HV and AJ conceptualized the manuscript. HV and FR conducted a literature review. HV, FR, and AJ drafted the manuscript. PV, RA and AJ provided comments and revised the manuscript. HV and FR created and designed all the figures. All authors read and approved the final version of the manuscript. It should be noted that AJ and RA are co-corresponding authors of this manuscript. Every author adhered to the COPE guidelines as universally acknowledged.